\documentclass[12pt, oneside]{article}   	
\usepackage{geometry} 
\usepackage{amsmath} 								
\usepackage{graphicx}	
\usepackage{amssymb}
\usepackage{cite}
\usepackage{color,xspace}
\usepackage{extarrows}
\usepackage{mathtools}

\usepackage[margin=1cm]{caption}

\numberwithin{equation}{section}
\DeclareSymbolFont{extraup}{U}{zavm}{m}{n}
\DeclareMathSymbol{\vardiamond}{\mathalpha}{extraup}{87}
\usepackage{float}
\restylefloat{table}
\allowdisplaybreaks

\def\twomat[#1,#2][#3,#4]{\left( \begin{array}{cc} #1 & #2 \\ #3 & #4 \end{array} \right)}
\def\thv[#1,#2,#3]{\left( \begin{array}{c} #1 \\ #2 \\ #3 \end{array} \right)}
\def\twv[#1,#2]{\left( \begin{array}{c} #1 \\ #2 \end{array} \right)}

\title{Revisiting the Scalar Weak Gravity Conjecture }

\date{}

\begin{document}

\begin{flushright}
\end{flushright}
\begin{center}

\vspace{1cm}
{\LARGE{\bf Revisiting the Scalar Weak Gravity Conjecture }}

\vspace{1cm}

\large{\bf Karim Benakli$^\spadesuit$ \let\thefootnote\relax\footnote{$^\spadesuit$kbenakli@lpthe.jussieu.fr},
Carlo Branchina$^{\vardiamond}$ \let\thefootnote\relax\footnote{$^\vardiamond$cbranchina@lpthe.jussieu.fr}
and
Ga\"etan~Lafforgue-Marmet$^\clubsuit$ \footnote{$^\clubsuit$glm@lpthe.jussieu.fr}
 \\[5mm]}

{ \sl Laboratoire de Physique Th\'eorique et Hautes Energies (LPTHE),\\ UMR 7589,
Sorbonne Universit\'e et CNRS, 4 place Jussieu, 75252 Paris Cedex 05, France.}

\end{center}
\vspace{0.7cm}

\abstract{ We revisit the Scalar Weak Gravity Conjecture and investigate the possibility to impose that scalar interactions dominate over gravitational ones. More precisely, we look for consequences of assuming that, for leading scalar interactions, the corresponding gravitational contribution is sub-dominant in the non-relativistic limit. For a single massive scalar particle, this leads us to compare four-point self-interactions in different type of potentials. For axion-like particles, we retrieve the result of the Axion Weak Gravity Conjecture: the decay constant $f$ is bounded by the Planck mass, $f < {M_{Pl}}$. Similar bounds are obtained for exponential potentials. For quartic, power law and Starobinsky potentials, we exclude large trans-Planckian field excursions.  We then discuss the case of moduli that determine the scalars  masses. We retrieve the exponential dependence as requested by the Swampland Distance Conjecture. We also find extremal state masses with field dependence that reproduces both the Kaluza-Klein and winding modes behaviour. In particular cases, our constraints can be put in the form of the Refined de Sitter Conjecture. 
}

\newpage
\setcounter{footnote}{0}

\section{Introduction}
\label{introduction}


Among the \textit{a priori} consistent low energy quantum field theories, it is believed that some cannot be embedded in a theory of quantum gravity. They form what is denoted as the {\it swampland }\cite{Vafa:2005ui,Ooguri:2006in} (see  \cite{Palti:2019pca, VafaCarta:2017} for a review). One of the selection criteria of consistent effective theories is provided by the Weak Gravity Conjecture (WGC) \cite{ArkaniHamed:2006dz}. It claims that, in a theory with  $U(1)$ gauge symmetry with coupling $g$, a state of charge $q$ and mass $m$ satisfying the inequality 
\begin{equation}
	g q \ge   \frac{ m}{{ M_{Pl}}}
	\label{WGCcharge-mass-ratio}
\end{equation}
must exist. Considering the charge over mass ratio, this condition can be obtained requiring that extremal black holes do decay entirely, leaving no remnants. It is furthermore consistent with black hole physics based arguments for non-existence of global symmetries in quantum gravity. This conjecture was claimed to be valid in any theory of quantum gravity and has been shown to hold in known examples in string theory.

There are two aspects of (\ref{WGCcharge-mass-ratio}) that are useful to stress. First, in theories with $\mathcal N \ge 2$ supersymmetries,  the central charge $Z$ of the supersymmetry algebra is given by $g q$ and is related to the mass of the BPS state through $|Z| = m$ (in Planck units). The relation (\ref{WGCcharge-mass-ratio}) goes in opposite direction of the BPS condition. It can be therefore tempting to look for other forms of such conjectures by considering the extremal states identities and turning it to an (anti-BPS) inequality. This was stressed in \cite{Palti:2017elp}.

The other useful aspect is the appearance of an ultraviolet scale $\Lambda\sim gq{ M_{Pl}}$, controlled by the gauge coupling $g$, which sets the cut-off of the EFT. This was dubbed as the magnetic weak gravity conjecture in \cite{ArkaniHamed:2006dz} and is clearly related to the non-existence of global symmetries in quantum gravity in the limit of weakly coupled gauge theories $g\to 0$.

Following the proposal of the WGC, another form was put forward as a Repulsive Force Conjecture (RFC)  \cite{Palti:2017elp,Heidenreich:2019zkl}. This postulates the existence of a state within the $U(1)$ theory with the property that, taken far apart, two copies of such state feel a repulsive force between each other. This avoids gravitational bound states. It was accurately described in \cite{Heidenreich:2019zkl}, where many of its consequences were exhibited.

Going beyond gauge fields and writing a similar conjecture for scalar fields, possibly complementary to swampland conjectures, is not straightforward. First, there is no such obvious argument on decay of black holes that can be used to induce the form of the conjecture. Second, to test in all generality different scalar conjectures in a quantum gravity theory is not easy. The scalar sector of the theory is very sensitive to the supersymmetry breaking. Implementing supersymmetry breaking in a string theory and extracting the full corrections to the scalar potential of a single real field in flat space-time is a non trivial problem. Moreover, supersymmetric models involve complex scalars, and it is not evident how to disentangle all the facets of constraints applying on one real scalar. With the lack of non-supersymmetric string theory examples, one is lead to postulate some form of the scalar conjecture and evaluate it by investigating the consequences. The hope is that even this modest trial and error method will turn out to be useful and will allow us to shed some light on the landscape of the effective field theories coupled to gravity. This way of proceeding applies to the conjectures discussed below.

A Scalar Weak Gravity Conjecture (SWGC) was investigated in \cite{Palti:2017elp} as a special case of the RFC. In the context of the RFC, the scalar field is massless and one is interested in the long range interactions it mediates. In an attempt to retrieve the Swampland Distance Conjecture mass formulae, it was proposed that: 
\begin{equation}
g^{ij}\partial _i m \partial_j m \ge  m^2
\label{Palti-SGC}
\end{equation}
where $\partial _i m \equiv \partial m /\partial \phi_i$ is the derivative of the mass term $m$ with respect to the scalar field $\phi_i$ and $g^{ij}$ is the appropriate metric on the space of fields.
In a footnote of \cite{Palti:2017elp}, it was also mentioned that, looking at different forms of the equalities satisfied by the central charge in $\mathcal N=2$, another possible form of the conjecture could have been:
\begin{equation}
	g^{ij}\partial _i \partial_j  m^2 \ge g^{ij}\partial _i m \partial_j m +  m^2.
	\label{Palti-SGC-2}
\end{equation}
The constraint (\ref{Palti-SGC}) does not involve repulsive interactions and as such cannot be considered as a realization of the RFC. It seemed puzzling in the RFC set-up discussed in \cite{Palti:2017elp}, as scalar mediated forces are attractive, and the  possibility (\ref{Palti-SGC-2}) was not pursued any further, with the exception of a few comments in \cite{DallAgata:2020ino}. It was somehow dismissed due to the lack of simple physical interpretation.

All these considerations led to the proposal of another form of the conjecture for scalar fields in \cite{Gonzalo:2019gjp}: the mass $m$ of an interacting scalar field satisfies the bound \cite{Ibanez:talk}:
\begin{equation}
	m^2 \frac {\partial^2}{\partial \phi^2}\left(\frac{1}{m^2}\right) \ge \frac{1}{{ M_{Pl}}^2}
	\label{Gonzalo-Ibanez-1}
\end{equation}
This was obtained by modifying by a factor 2 and  an additional four-point contact interaction the inequality (\ref{Palti-SGC}) expressed as derivatives of the scalar potential. This form of the conjecture was motivated by a set of implications\cite{Gonzalo:2019gjp,Ibanez:talk,Kusenko:2019kcu,Shirai:2019tgr,Andriot:2020lea}, some of which might be of phenomenological importance. However, it raises some questions about its origin and the meaning of the corresponding inequality. As a consequence of the (\ref{Gonzalo-Ibanez-1}), for states with a mass depending on the scalar $\phi$, the equality in (\ref{Gonzalo-Ibanez-1}) is reached for
\begin{equation}
m^2 (\phi) = \frac{m_0^2}{ A e^{-\phi}+ B e^{\phi}}
\label{Gonzalo-Ibanez-2}
\end{equation}
where $A$ and $B$ are integration constants. Through the identification $e^{-\phi}=R^2$, the result of (\ref{Gonzalo-Ibanez-2}) has been interpreted in \cite{Gonzalo:2019gjp} as an indication of the extended nature of the fundamental states.

Taken as such, the above proposals were dismissed in \cite{Freivogel:2019mtr}, because of inconsistent implications for simple scalar potentials, and it was instead suggested that scalar particles should be subject to constraints in such a way that they would not form bound states with size smaller than their Compton wavelength. No generic alternative formulation for these constraints on the scalar potential was proposed.

In this work, we will postulate that in the appropriate low energy limit, for the leading interaction, the gravitational contribution must be sub-leading. For particular scalar fields, we will propose an explicit set-up, based on the computation of four-point functions, for comparing the different interactions.  The resulting  inequalities will reproduce different forms of the Swampland conjectures, and, in a particular case, the inequality will be saturated for masses of the form (\ref{KK-winding}):
\begin{equation}
 m_X^2(\phi)= m_{-}^2 e^{-{2}\phi} + m_{+}^2 e^{{2}\phi} \, .
\label{KK-winding-predict}
\end{equation}
instead of (\ref{Gonzalo-Ibanez-2}).

The paper is organized as follows. In Section 2, we formulate the constraint of dominance of scalar interactions with respect to the gravitational ones for the case of a single massive scalar field self-interacting. We illustrate the constraint by the simplest example of a single real field with a cubic and quartic potential. A few other examples are studied in  section 3. Those include the quartic complex potential, the axion, the exponential and the Starobinsky potential. In the section 4, we discuss an extension to moduli and massless scalars. Section 5 presents our conclusions.

%
%

\section{Scalar vs Gravity in the Non-Relativistic Regime}

The Weak Gravity Conjecture states that for any abelian gauge symmetry $U(1)$ there is at least one state with gauge self-interaction stronger than the gravitational one. Here, we will investigate a possible extension of the conjecture to the case of scalar fields. 

We start with the case of a single self-interacting {\it massive} scalar field.  We will postulate that for this scalar field the self-interaction is stronger than the gravitational one.

This assertion calls for a few immediate remarks. First, we need to specify at which scale the different interactions are computed and compared. This is chosen to be of order of the mass of the self-interacting particle. This is consistent with the fact that the Weak Gravity Conjecture makes statements about properties of effective field theories. At these energy scales, the non-relativistic theory is a good approximation. This means, for example, that in scattering processes the particle number is conserved. We shall therefore investigate the strength of the interactions by computing the simplest scattering processes. Precisely, we will compare the four-point amplitude contribution of the scalar self-interaction versus the gravitational one.

We work in the non-relativistic limit and keep only the leading order in $1/c^2$. The gravitational forces are then expected to be well described by the Newtonian potential. Higher order corrections, as those given by the Einstein–Infeld–Hoffman Lagrangian, will be neglected. In practice, instead of dealing with the potential in coordinates space, we will work in the Fourier-transform space by computing the scattering amplitudes. The dominance of scalar self-interaction means in particular that all the higher dimensional non-renormalizable interactions suppressed by higher powers of the Planck mass should be subdominant and may be neglected. We will see below that this preeminence can happen to be violated in isolated regions of size $\frac{\Delta \phi^2}{m^2} \sim \frac{m^2}{{\tilde M_{Pl}}^2}$ where the interactions can switch nature between attractive and repulsive.

\begin{figure}
\centering
\includegraphics[scale=0.28]{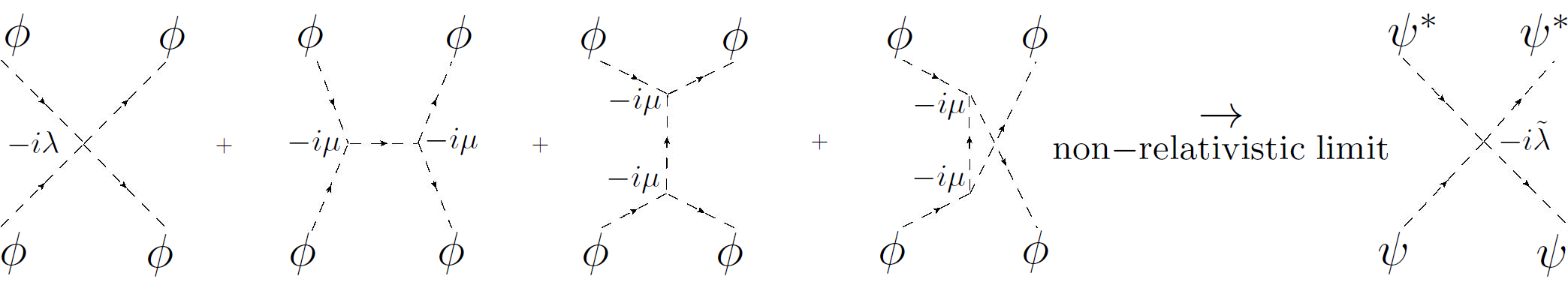}
\caption{The identification of the $2\rightarrow2$ scattering in the non relativistic theory coming from the corresponding scattering in the relativistic case.}
\label{diagrams}
\end{figure}

We restrict to four-dimensional Minkowski space-time and use from now on natural units $\hbar = c =1$. 
We first investigate the simplest case of cubic and quartic potential and discuss other forms of scalar potentials in the next section.
\\
We consider a real scalar $\phi$ with the potential:
\begin{equation}
	V(\phi)=\frac{1}{2}m_0^2\phi^2+\frac{\mu}{3!}\phi^3+\frac{\lambda}{4!}\phi^4.
	\label{potential real scalar}
\end{equation}
In string theory, our fiducial quantum gravity theory, all the low energy parameters are field dependent. But we will consider here that the other scalar fields are fixed to their vacuum value and decouple from the dynamics of the low energy effective action under scrutiny.  At energy scales $E \sim m_0$, the theory is non-relativistic and can be described by the corresponding limit. We study fluctuations around $\phi=0$ and make the field redefinition:
\begin{equation}
	\phi (x)   =\frac{1}{\sqrt{2m_0}}\left(\psi(\mathbf{x},t)e^{-im_0t}+\psi^{*}(\mathbf{x},t)e^{im_0t}\right)
	\label{psi def}
\end{equation}
where the phase $e^{-im_0t}$ is introduced to take into account the leading $m_0$ term in the non-relativistic limit expansion $E \simeq m_0+ {\mathbf{p^2}} /2m_0$ where ${\mathbf{p}}$ is the particle three-dimensional momentum. The denominator ${\sqrt{2m_0}}$ comes from the different normalizations in relativistic and non-relativistic quantum mechanics.
\\
The potential for the non-relativistic field $\psi$ should be of the form

\begin{equation}
	V_{eff}\left(\psi\psi^{*}\right)=m_0\psi\psi^{*}+\frac{\tilde{\lambda}}{16m_0^2}\left(\psi\psi^{*}\right)^2.
\end{equation}
We now want to relate the single non-relativistic coupling $\tilde{\lambda}$ with the coefficients of the relativistic potential. We identify the low energy limit of the $2\rightarrow2$ scattering in the $\phi$ description with the corresponding scattering of four $\psi$ states. This leads trivially to $\lambda=\tilde{\lambda}$ when $\mu=0$ in \eqref{potential real scalar}. In the case where $\mu \neq 0$, we will have to take into account the contributions to the $2\rightarrow2$ scattering from the exchange of a virtual $\phi$. We have in this case three diagrams, one for each channel, as shown in Figure \ref{diagrams}. We can compute the non relativistic limit of each one of them. This is obtained requiring $s-4m_0^2 \ll m_0^2$, where $s= (p_1+p_2)^2$ is the usual Mandelstam variable and $p_1$, $p_2$ the four-momenta of the initial states. We also have $t=-\frac{1}{2}(s-4m_0^2)(1-\mathrm{cos}(\theta))$ and $u=-\frac{1}{2}(s-4m_0^2)(1+\mathrm{cos}(\theta))$, $\theta$ being the angle between the in-going and out-going particles momenta in the center of mass frame. This basic computation yields the s-channel contribution as:  
\begin{equation}
	\left(-i\mu\right)^2\frac{i}{s-m_0^2}=\frac{-i\mu^2}{3m_0^2}+{\mathcal O}\left(\frac{s-4m_0^2}{m_0^2}\right),
\end{equation}
and the t-channel as:
\begin{equation}
	\left(-i\mu\right)^2\frac{i}{t-m_0^2}=\frac{i\mu^2}{m_0^2}+{\mathcal O}\left(\frac{s-4m_0^2}{m_0^2}\right).
\end{equation}
Finally, the u-channel contribution is the same as the t-channel one. Summing up the three contributions we obtain $i\frac{5}{3}\frac{\mu^2}{m_0^2}$, so that the effective four-point self-interaction coupling in the non-relativistic limit is:

\begin{equation}
	\label{lambdatilde}
	\tilde{\lambda}=\lambda-\frac{5}{3}\frac{\mu^2}{m_0^2}.
\end{equation}

In computing the gravitational interaction, we have  assumed $m_0^2>0$. Both attractive and repulsive forces can be obtained from the quartic self-interaction, through the choice of $\lambda<0$ and $\lambda>0$ respectively. On the other hand, the trilinear term always leads to an attractive force in a $2\to2$ states scattering. However, when $\lambda<0$ the stability of the potential means that additional non renormalizable terms are important and should be taken into account.
In the case of $\lambda>0$, eq (\ref{lambdatilde}) shows  the competition between the attractive and repulsive interactions in the non-relativistic limit. The resulting sign of $\tilde{\lambda}$ tells us about the attractive or repulsive nature of the effective interaction and, in the case where they are in competition, which one of the two terms dominate at energies $E\sim m_0$. 

In the WGC the gauge and gravity forces have similar dependence in the distance between the scattering particles at leading order. There are two corrections, one from the evolution of gauge coupling with energy and the other from post-Newtonian effects. This is not the case for the scalar interaction. In the non-relativistic limit, the scalar potential is approximated by a delta distribution in space while the gravitational potential is Newtonian. A point-like interaction arises from integrating out massive mediators. In the infrared, at energies below the mass scale, the gravitational scattering exhibits a divergence coming from the $t$ and $u$ channels. Obviously, to compare a Newtonian potential at long distance with the strength of the scalar localised interaction is not very instructive. It is essential in the comparison to fix the energy scale, and naturally it is given by the mass of the scalar particle, and consider the gravitational scattering in the s-channel at $s \sim 4 m^2_0$.

Requiring that gravity is the weakest force at low energy amounts then to impose:
\begin{equation}
	\left|\tilde{\lambda}\right|=\left|\lambda-\frac{5}{3} \frac{\mu^2}{m_0^2}\right|\ge \frac{m_0^2}{M_{Pl}^2}.
	\label{ourSWGC}
\end{equation}
We have put an absolute value on the left hand side so that it holds independently of the sign of the self-interaction. Note also that, in the spirit of \cite{ArkaniHamed:2006dz,Lust:2017wrl,Craig:2019fdy}, the quantity $\sqrt{|\tilde\lambda|} M_{Pl}$, could be interpreted as an ultra-violet cut-off scale dictated by quantum gravity. In particular, this means that both the limits $\lambda \to 0$ and $\mu \to 0$ cannot be taken simultaneously. Cancellation of the two terms in $\tilde\lambda$, as we said, might encode the change of nature of the scalar interactions on a region of the phase space that need to be studied case by case.

Below, we will work in more generic field background values and potentials, therefore we will impose a stronger condition
\begin{equation}
4 m_0^2 \bigg\lvert \frac {\partial^4 V_{eff}}{\partial^2 \psi {\partial}^2{ \psi^*}} \bigg\rvert_{\psi = 0}  \ge  \frac {\tilde c}{M_{Pl}^2} \bigg\lvert  \frac {\partial^2 V_{eff}}{\partial \psi {\partial}{\psi^*}} \bigg\rvert^2_{\psi = 0} 
\label{conjecture}
\end{equation}
and take the order one constant ${\tilde c}$ to be ${\tilde c} =1$, which amounts to redefine the Planck mass to ${\tilde M_{Pl}}$. The r.h.s. of (\ref{conjecture}) represents the gravitational attractive interaction between the two particles only when we work at the minimum of the potential and the squared mass is positive defined.

We focus now on the simplest case $\mu =0$ and investigate the relative 	strengths of self-interaction and gravitational one when $\phi $ sweeps the range of possible values. For this purpose we  consider small perturbations $\delta \phi$, corresponding to the above $\psi$, around background values $\phi$. We expand	
\begin{equation}
	V(\phi+\delta\phi)=\frac{1}{2}m_0^2\phi^2+\frac{1}{4!}\lambda\phi^4+m_0^2\phi\delta\phi+\frac{\lambda}{3!}\phi^3\delta\phi+\frac{1}{2}\left(m_0^2+\frac{\lambda}{2}\phi^2\right)(\delta\phi)^2+\frac{\lambda}{3!}\phi(\delta\phi)^3+\frac{\lambda}{4!}(\delta\phi)^4.
	\label{quarticpotential}
\end{equation}

From (\ref{quarticpotential}), we can immediately read the mass term, the cubic and the quartic couplings for  $\delta\phi$ and the effective quartic coupling in the non-relativistic limit. Those are given by:
\begin{eqnarray}
	m^2_{\delta\phi}(\phi)=m_0^2+\frac{\lambda}{2}\phi^2, \qquad \mu_{\delta\phi}=\lambda\phi, \qquad \lambda_{\delta\phi}=\lambda \qquad \tilde\lambda=\lambda-\frac{5}{3}\frac{\lambda^2\phi^2}{m_0^2+\lambda/2\phi^2}.
	\label{non-relativistic-parameters-quarticpotential}
\end{eqnarray}

  We restrict to the case with $m_0^2, \lambda>0$ to explicitly exhibit the competition between the attractive and repulsive terms. Requiring gravity to be the weakest force leads to
\begin{equation}
	\left|\lambda-\frac{5}{3}\frac{\lambda^2\phi^2}{m_0^2+\frac{\lambda}{2}\phi^2}\right|\ge\frac{1}{{\tilde M_{Pl}}^2}\left(m_0^2+\frac{\lambda}{2}\phi^2\right).
	\label{Quarticnon-relativisticinequality}
\end{equation}
The term inside the absolute value of (\ref{Quarticnon-relativisticinequality}) vanishes for $\phi^2=\frac{6}{7}\frac{m_0^2}{\lambda}$. The cubic term dominates above this turning point, a region where the interaction is attractive. The quartic one dominates instead below the turning point, making the scalar interaction repulsive. 	

We first investigate the $\phi^2\le\frac{6}{7}\frac{m_0^2}{\lambda}$ region where (\ref{Quarticnon-relativisticinequality}) reads 
\begin{equation}
	\phi^4+\left(4\frac{m_0^2}{\lambda}+\frac{14}{3}{\tilde M_{Pl}}^2\right)\phi^2+4\frac{m_0^4}{\lambda^2}-4{\tilde M_{Pl}}^2\frac{m_0^2}{\lambda}\le 0.
\label{quarticinequalityfirstcase}
\end{equation}  
Assuming $\lambda\ge\frac{m_0^2}{{\tilde M_{Pl}}^2}$ and discarding the solutions with $\phi^2<0$, this is verified inside the region
\begin{equation}
	0\le\phi^2\le-2\frac{m_0^2}{\lambda}-\frac{7}{3}{\tilde M_{Pl}}^2+\frac{7}{3}{\tilde M_{Pl}}^2\sqrt{1+\frac{120}{49}\frac{m_0^2}{\lambda {\tilde M_{Pl}}^2}}.
	\label{region1quartic}
\end{equation}
At the first order in $\frac{1}{{\tilde M_{Pl}}^2}$, this is obtained:
\begin{equation}
	\phi^2 \lesssim \frac{6}{7} \frac{m_0^2}{\lambda} - \frac{600}{343} \frac{m_0^4}{\lambda^2} \frac{1}{{\tilde M_{Pl}}^2}
\end{equation}
which exhibits a small region of order ${\tilde M_{Pl}}^{-2}$ below the critical value where gravity is stronger than quartic scalar self-interaction.

For $\lambda\le\frac{m_0^2}{{\tilde M_{Pl}}^2}$, the turning point happens at a scale $\phi^2\sim\frac{m_0^2}{\lambda}\ge {\tilde M_{Pl}}^2$ and, as the inequality would not be solved for $\phi^2\le\frac{6}{7}\frac{m_0^2}{\lambda}$, this would translate in gravity being stronger than scalar interactions all the way up to the Planck scale.

Let us now turn to the case $\phi^2\ge\frac{6}{7}\frac{m_0^2}{\lambda}$. There, the inequality translates into 
\begin{equation}
	\phi^4+\left(4\frac{m_0^2}{\lambda}-\frac{14}{3}{\tilde M_{Pl}}^2\right)\phi^2+4\frac{m_0^4}{\lambda^2}+4{\tilde M_{Pl}}^2\frac{m_0^2}{\lambda}\le 0.
\end{equation}  
At leading order in $\frac{m_0^2}{{\tilde M_{Pl}}^2}$, the region where the inequality is verified is given by
\begin{equation}
	\frac{6}{7} \frac{m_0^2}{\lambda} + \frac{600}{343} \frac{m_0^2}{\lambda^2} \frac{m_0^2}{{\tilde M_{Pl}}^2} \lesssim \phi^2 \lesssim \frac{14}{3}{\tilde M_{Pl}}^2 - \frac{6}{7}\frac{m_0^2}{\lambda} +\mathcal{O}({\tilde M_{Pl}}^{-2})
	\label{region2quartic}
\end{equation}
In conclusion, up to the Planck scale, the gravity seems to dominate only around the special value $\phi^2=\frac{6}{7}\frac{m_0^2}{\lambda}$ in a symmetric interval of radius ${\Delta \phi^2} \sim \frac{m_0^4}{{\tilde M_{Pl}}^2}$. It would be interesting to investigate, for explicit examples of quantum gravity, if the theory can be insensitive to such small field excursion regions, but this goes beyond the scope of this work.


\section{Single Scalar Field Potentials}	
	
In this section, we would like to investigate what the implications of requiring gravity to be weaker than the scalar field self-interactions in the non-relativistic limit are on different potentials. More precisely, we will consider {\it very slowly rolling fields}, having in mind possible cosmological applications. We  impose the condition (\ref{conjecture}) and extract its implications for the involved scales and couplings.



\subsection{The Mexican Hat or Higgs-like Quartic Potential}

We consider the quartic scalar potential
    	\begin{equation}
V(\phi,{\bar \phi})= - m^2 {\bar \phi} \phi+\lambda({\bar \phi} \phi)^2.    
\label{Higgspotential}
\end{equation}
with $\lambda > 0$, insuring stability, and $m^2 >0$.

 It is convenient to use the parametrization $\phi(x)=\frac{1}{\sqrt 2}\rho(x)e^{i\pi(x)}$.  This potential develops a minimum at  $\rho^2 = \frac{m^2}{\lambda}$. This theory has 
 a global $U(1)$ symmetry {\footnote{ Quantum gravity requires that either the symmetry is gauged or  broken. However, the latter might be sub-leading to the quartic self-interaction considered here.}}, which is spontaneously broken at the minimum, and $\pi(x)$ is the associated Goldstone boson. The final mass of $\pi(x)$ depends on details of the complete theory. It might be generated by the higher order terms breaking the global symmetry, as dictated for instance by the WGC. It could also be that the $U(1)$ symmetry is gauged. Then $\pi(x)$ gives rise to the longitudinal mode of the massive gauge boson. We will focus here only on the field $\rho (x)$ which plays in the latter case the role of the Higgs field.

We consider a small perturbation $\delta\rho(x)$ around a the background value $\rho(x)$. The expansion of the potential, up to $\mathcal{O}(\delta\rho^4)$, reads:
	
\begin{equation}
	V(\rho+\delta\rho) \simeq -\frac{1}{2}m^2\rho^2 + \frac{\lambda}{4}\rho^4 + (\lambda\rho^3-m^2\rho)\delta\rho + \frac{1}{2}(3\lambda\rho^2-m^2)\delta\rho^2 + \lambda\rho\delta\rho^3 + \frac{\lambda}{4}\delta\rho^4.
\end{equation}
The effective mass term, trilinear and quartic couplings of $\delta\rho(x)$ are then given by $m^2_{\delta\rho}=3\lambda\rho^2-m^2$, $\mu_{\delta\rho}=6\lambda\rho$, $\lambda_{\delta\rho}=6\lambda$, respectively. The  $\delta\rho(x)$  resulting quartic self-interaction $\tilde{\lambda}$ at low energies can now be computed to be
	
\begin{equation}
\tilde{\lambda}=6\lambda-\frac{60\lambda^2\rho^2}{3\lambda\rho^2-m^2}= - 6 \lambda \frac{ (m^2 +7 \lambda \rho^2)}{ 3\lambda\rho^2 - m^2 }.    
\end{equation}
Vanishing self-interaction, i.e. a null value for $\tilde{\lambda}$, corresponds to $ m^2\lambda+7\lambda^2\rho^2=0$. This is obviously never satisfied here. 

We discard the region $\rho^2<\frac{m^2}{3\lambda}$ where the effective mass of $\delta\rho(x)$  is either tachyonic or vanishing, though we have checked that the inequality (\ref{conjecture}) is satisfied.

We will investigate the region $m^2_{\delta\rho}>0$, i.e. $\rho^2>\frac{m^2}{3\lambda}$. We have:
\begin{equation}
9 \frac{\lambda^2}{{\tilde M_{Pl}}^2} \rho^4 - \left(6\lambda\frac{m^2}{{\tilde M_{Pl}}^2}+42\lambda^2\right)\rho^2+\frac{m^4}{{\tilde M_{Pl}}^2}-6m^2\lambda\le0.
\end{equation}

Discarding the region $\rho^2 \in \left[0,\frac{m^2}{3\lambda} \right]$  as discussed above, the inequality is satisfied for:

\begin{equation}
\frac{m^2}{3\lambda} <  \rho^2  \leqslant \frac{14}{3} {\tilde M_{Pl}}^2 + \frac{17}{21} \frac{m^2}{\lambda} + \mathcal O({\tilde M_{Pl}}^{-2}) 
\end{equation}
It is worth mentioning that at the minimum, where $\rho^2=\frac{m^2}{\lambda}\equiv v$, we get $\tilde{\lambda}=-24\lambda$, and the conjecture is then verified in the case:
	
\begin{equation}
\label{Higgsconstraint}
\lambda\ge\frac{1}{12}\frac{m^2}{{\tilde M_{Pl}}^2}\sim10^{-17}\Leftrightarrow v^2\le12{\tilde M_{Pl}}^2\sim10^{37}GeV^2,   
\end{equation}
where we have taken $m$ to be the electroweak scale.



\subsection{Axion-like Potential}
\label{subsection-Axion}

Let's consider the case of the axion potential:
\begin{equation}
	V(\phi)=\mu^4\left(1-\cos\left(\frac{\phi}{f_a}\right)\right).
	\label{Axionpotential}
\end{equation}
Expanding this potential around a fixed value $\phi_0$ and excluding points where $\cos\left(\frac{\phi_0}{f_a}\right)=0$ as the state becomes massless and our non-relativistic limit no more applies, we obtain up to fourth order  in $\delta \phi$:

\begin{eqnarray}
	V(\phi) \simeq &\mu^4\left[1-\cos\left(\frac{\phi_0}{f_a}\right)+\sin\left(\frac{\phi_0}{f_a}\right)\frac{\delta\phi}{f_a}+\frac{1}{2}\cos\left(\frac{\phi_0}{f_a}\right)\frac{(\delta\phi)^2}{f_a^2} \right.\nonumber \\
	&\left. -\frac{1}{3!}\sin\left(\frac{\phi_0}{f_a}\right)\frac{(\delta\phi)^3}{f_a^3}-\frac{1}{4!}\cos\left(\frac{\phi_0}{f_a}\right)\frac{(\delta\phi)^4}{f_a^4}\right],
\end{eqnarray}
from which we can read  $\tilde{\lambda}=-\frac{1}{f_a^4}\left(\cos\left(\frac{\phi_0}{f_a}\right)+\frac{5}{3}\frac{\sin^2(\phi_0/f_a)}{\cos(\phi_0/f_a)}\right)$. Requiring gravity to be the weakest force leads to 
\begin{equation}
\frac{1}{f_a^2}\left|\cos\left(\frac{\phi_0}{f_a}\right)+\frac{5}{3}\frac{\sin^2\left(\frac{\phi_0}{f_a}\right)}{\cos\left(\frac{\phi_0}{f_a}\right)}\right| \ge \frac {1}{{\tilde M_{Pl}}^2}{\left|\cos\left(\frac{\phi_0}{f_a}\right)\right|},
\label{axioninequality}
\end{equation}
which yields 
\begin{equation}
\frac{1}{f_a^2}\left|1+\frac{5}{3}\tan^2\left(\frac{\phi_0}{f_a}\right)\right|\ge\frac{1}{{\tilde M_{Pl}}^2}.
\label{WGC axion case}
\end{equation}
We have expanded around a generic background value $\phi_0$ thus (\ref{WGC axion case}) leads to:
\begin{equation}
f_a^2\le {\tilde M_{Pl}}^2  
\end{equation}
We therefore retrieve the Axion Weak Gravity Conjecture, which requires an axion decay constant lower the Planck scale \cite{ArkaniHamed:2006dz,Rudelius:2015xta,Montero:2015ofa,Brown:2015iha,Heidenreich:2015wga,delaFuente:2014aca,Hebecker:2015rya,Bachlechner:2015qja,Rudelius:2014wla,Junghans:2015hba,Kooner:2015rza,Ibanez:2015fcv,Hebecker:2015zss,Hebecker:2019vyf,Daus:2020vtf}. Note that, in the r.h.s. of (\ref{axioninequality}), we have taken the absolute value of the squared mass term. Here we see the inequality as taken on derivatives of the potential since the squared mass can be negative.

\subsection{Inverse power-law effective scalar potential}

Another scalar potential is the inverse power-law one, frequently used in cosmological applications. It reads
\begin{equation}
V(\phi)=M^{4+p}\phi^{-p},
\label{Inverseplpotential}
\end{equation}
where $p>0$ is a constant and $M$ sets the energy scale.
In the general case, we expand the potential as a Taylor series
\begin{eqnarray}
\frac{1}{M^{4+p}}V(\phi_0+\delta\phi)&=&\phi_0^{-p}-p\phi_0^{-p-1}\delta\phi+\frac{p(p+1)}{2}\phi_0^{-p-2}(\delta\phi)^2-\frac{p(p+1)(p+2)}{3!}\phi_0^{-p-3}(\delta\phi)^3\nonumber \\&+&\frac{p(p+1)(p+2)(p+3)}{4!}\phi_0^{-p-4}(\delta\phi)^4.
\end{eqnarray} 
The effective quartic interaction in the non-relativistic limit is given by 

\begin{align}
\tilde{\lambda}=-\frac{p(p+1)(p+2)}{3}(2p+1)\phi_0^{-p-4} \, .
\end{align}
The gravitational interaction will thus be weaker than the scalar self-interaction in the non-relativistic limit if
\begin{equation}
	\frac{p(p+1)(p+2)}{3}(2p+1)\left|\phi_0^{-p-4}\right|\ge\frac{p(p+1)}{{\tilde M_{Pl}}^2}\left|\phi_0^{-p-2}\right|.
\end{equation}
which  is satisfied for 
\begin{equation}
\phi_0^2\le\frac{(p+2)(2p+1)}{3}{\tilde M_{Pl}}^2 ,   
\end{equation}
therefore forbidding large trans-Planckian excursions.

%
%

\subsection{Exponential Scalar Potential}

Another popular class of scalar potentials is represented by sums of exponential functions. We focus here on the simplest case 
\begin{equation}
 V(\phi)=\Lambda_0 e^{-\lambda\phi/f}.
\label{Exponentialpotential}
\end{equation}
The expansion around a background value $\phi_0$ reads
\begin{eqnarray}
V(\phi_0+\delta\phi)= \Lambda_0 e^{-\lambda\phi_0/f}\left[1-\lambda\frac{\delta\phi}{f}+\frac{1}{2}\lambda^2\left(\frac{\delta\phi}{f}\right)^2-\frac{1}{3!}\lambda^3\left(\frac{\delta\phi}{f}\right)^3+\frac{1}{4!}\lambda^4\left(\frac{\delta\phi}{f}\right)^4\right],
\label{Exponentialdevelopment}
\end{eqnarray}
and the self-interaction of the scalar field in the non-relativistic limit is encoded in the $\tilde{\lambda}$ quartic coupling
\begin{equation}
\tilde{\lambda}=\Lambda_0 e^{-\lambda\phi_0/f}\left(\frac{\lambda^4}{f^4}-\frac{5}{3}\frac{\lambda^4}{f^4}\right)=-  \frac{2}{3}\frac{\lambda^4}{f^4}\Lambda_0 e^{-\lambda\phi_0/f}.
\end{equation}
Application of our bound is straightforward and yields the following inequality

\begin{equation}
\frac{2}{3}\frac{\lambda^2}{f^2}\ge\frac{1}{{\tilde M_{Pl}}^2},
\end{equation}
The weak gravity regime under scrutiny is realized for scalars with an exponential potential as long as their scale does not exceed the Planck one, with 
\begin{equation}
f^2 \le \frac{2}{3} \lambda^2 {\tilde M_{Pl}}^2 .   
\end{equation}
This bound still allows for a cosmological expansion (see e.g. \cite{Tsujikawa:2013fta}), but is in conflict with the requirement obtained in \cite{Agrawal:2018own}, as we will discuss below.

Let's consider the case of a double exponential potential 
\begin{equation}
\label{Doubleexponential}
	V(\phi)=\Lambda_1e^{-\lambda_1\phi/f}+\Lambda_2e^{-\lambda_2\phi/f},
\end{equation}
with the assumption $\lambda_1 \sim \lambda_2$. We develop each exponential as in (\ref{Exponentialdevelopment}) to get
\begin{equation}
\tilde{\lambda}=\Lambda_1\frac{\lambda_1^4}{f^4}e^{-\lambda_1\phi_0/f}+\Lambda_2\frac{\lambda_2^4}{f^4}e^{-\lambda_2\phi_0/f}-\frac{5}{3}\frac{1}{f^4}\frac{\left(\Lambda_1{\lambda_1^3} e^{-\lambda_1\phi_0/f}+\Lambda_2 {\lambda_2^3} e^{-\lambda_2\phi_0/f}\right)^2}{\Lambda_1 {\lambda_1^2} e^{-\lambda_1\phi_0/f}+\Lambda_2{\lambda_2^2} e^{-\lambda_2\phi_0/f}},
\end{equation}
which can be rewritten as 

\begin{equation}
	\tilde{\lambda}=- \frac{1}{f^4} \frac{\frac{2}{3}\Lambda_1^2\lambda_1^6 e^{-2\lambda_1\phi_0/f} 
	+ \frac{2}{3}\Lambda_2^2\lambda_2^6e^{-2\lambda_2\phi_0/f} +\Lambda_1\Lambda_2\lambda_1^2\lambda_2^2\left(\frac{10}{3}\lambda_1\lambda_2 -\lambda_1^2-\lambda_2^2\right)e^{-(\lambda_1+\lambda_2)\phi_0/f}}
	{\Lambda_1 {\lambda_1^2} e^{-\lambda_1\phi_0/f}+\Lambda_2 {\lambda_2^2} e^{-\lambda_2\phi_0/f}}. 
\end{equation}
The analysis of this constraint on a double exponential is somehow quite involved, and not useful here to discuss in full generality. In the case where $\lambda_1^2+\lambda_2^2\le\frac{10}{3}\lambda_1\lambda_2$, all three terms in the numerator have the same sign. For $\Lambda_{1,2}>0$ ($\Lambda_{1,2}<0$) the scalar self-interaction is attractive (repulsive). The condition for gravity to be the weakest force reads

\begin{eqnarray}
I(\Lambda_1,\Lambda_2,\lambda_1, \lambda_2,f) & = & 	\lambda_1^4 \Lambda_1^2\left(\frac{2}{3}\frac{\lambda_1^2}{f^2}-\frac{1}{{\tilde M_{Pl}}^2}\right)e^{-2\lambda_1\phi_0/f}
	 +\lambda_2^4 \Lambda_2^2\left(\frac{2}{3}\frac{\lambda_2^2}{f^2}-\frac{1}{{\tilde M_{Pl}}^2}\right)e^{-2\lambda_2\phi_0/f} \nonumber \\
	& & +\Lambda_1\Lambda_2\lambda_1^2\lambda_2^2\left(\frac{10/3\lambda_1\lambda_2-\lambda_1^2-\lambda_2^2}{f^2}-\frac{2}{{\tilde M_{Pl}}^2}\right)e^{-(\lambda_1+\lambda_2)\phi_0/f} \nonumber \\ 
	&\ge 0&
\end{eqnarray}
It is verified for mass scales not exceeding the value $f^2\sim\frac{2}{3}\lambda_{1,2}^2{\tilde M_{Pl}}^2$.

%
%

\subsection{Starobinsky Potential}

The power-law and the exponential potentials are frequently used in early Universe cosmology. We investigate here the implications of (\ref{conjecture}) for the Starobinsky's potential \cite{Starobinsky:1980te}.

We consider the potential: 

\begin{equation}
	V(\phi)=\Lambda^4\left(1-e^{-\sqrt{2/3}\phi/{\tilde M_{Pl}}}\right)^2
\end{equation}
and expand it around a background field value $\phi_0$, and study the leading order contribution to the quartic self-interaction perturbation $\delta\phi = \phi -\phi_0$. The non-relativistic regime quartic coupling $\tilde{\lambda}$ is given by:
\begin{eqnarray}
\frac{{\tilde M_{Pl}}^4}{\Lambda^4}\tilde{\lambda}&=&\frac{-\frac{256}{27}e^{-4\sqrt{2/3}\phi_0/{\tilde M_{Pl}}}+\frac{80}{27}e^{-3\sqrt{2/3}\phi_0/{\tilde M_{Pl}}}-\frac{16}{27}e^{-2\sqrt{2/3}\phi_0/{\tilde M_{Pl}}}}{{2}e^{-2\sqrt{2/3}\phi_0/{\tilde M_{Pl}}}-e^{-\sqrt{2/3}\phi_0/{\tilde M_{Pl}}}}.
\end{eqnarray}
The weakness of the gravitational interaction reads now
\begin{equation}
	\left|-\frac{16}{9}e^{-2\sqrt{2/3}\phi_0/{\tilde M_{Pl}}}+\frac{5}{9}e^{-\sqrt{2/3}\phi_0/{\tilde M_{Pl}}}-\frac{1}{9}\right|\ge \left|e^{-2\sqrt{2/3}\phi_0/{\tilde M_{Pl}}}-e^{-\sqrt{2/3}\phi_0/{\tilde M_{Pl}}}+\frac{1}{4}\right|,
	\label{Starobinskynon-relativisticconstraint}
\end{equation}
where we have put the absolute value on the r.h.s. to stress its positivity even if it is useless, being the square of a real quantity. Nevertheless, we still should study the sign and the strength of the l.h.s. of (\ref{Starobinskynon-relativisticconstraint}). The term inside the absolute value is always negative, meaning the scalar interaction is always attractive. So we can just drop the absolute values in eq. (\ref{Starobinskynon-relativisticconstraint}). Simple algebra finally leads us to the conclusion that gravity is weaker than the scalar self-interaction if 

\begin{equation}
 \phi_0\le\sqrt{\frac{3}{2}}\ln\left(\frac{14}{\sqrt{51}-4}\right){\tilde M_{Pl}} \sim 2 {\tilde M_{Pl}}.
\end{equation}    
The coefficient in front of ${\tilde M_{Pl}}$ in the above equation is of order $1$. Slitghly before reaching this scale, we would encounter tachyonic modes for $\phi\sim\sqrt{\frac{3}{2}}\ln{(2)}\tilde M_{Pl}$. In this Starobinsky's model, self-interactions are strong enough to keep gravity the weakest force all the way up to the Planck scale.

%

%

%
\subsection{Weak Gravity and Quintessence}

One of the popular use of the above scalar potentials is for inducing cosmic acceleration, more precisely using $\phi$ as the quintessence field. We discuss here some direct implications of our constraints for such applications.

The late time cosmic acceleration may indeed be understood either in terms of a cosmological constant, in the context of the $\Lambda$CDM, or in terms of a dynamical scalar field, slowly rolling towards the minimum of its potential \cite{Ratra:1987rm,Wetterich:1987fm}. In the equation of state, the ratio pressure/energy density $w$ is fixed to the value $\mathit{w}=-1$ in the first case, while it becomes a dynamical variable in the case of the quintessence \cite{Caldwell:1997ii}. The swampland criteria seem to be in favor of the latter scenario, that, with parameters tightened by the current observations, may fit into the program (see \cite{Agrawal:2018own}). In this context, 
for the dark energy to take over the control of the expansion of the Universe at late times, the quintessence field needs to be very light, with mass of order the Hubble parameter as measured today $m\lesssim H_0\sim10^{-33}eV$. The corresponding potential is unknown and forms similar to those studied above have been considered (see for a review \cite{Tsujikawa:2013fta}). Requirements for the evolution equations of a scalar field $\phi$ to have a fixed point realizing the desired equation of state can be expressed as

\begin{equation}
\begin{cases}
\mathit{w}_{eff}\equiv\frac{\rho_\phi+\rho_m}{P_\phi+P_m}=\mathit{w}_\phi>-\frac{1}{3}; \\	
\Omega_\phi\equiv\frac{\rho_\phi}{3M_{Pl}^2H^2}=1,
\end{cases}
\end{equation}
where we denote with the subscript $m$ the matter contribution, and $\phi$ for the quintessence one, and\cite{Copeland:1997et}:

\begin{equation}
\label{Quintessencecondition}
\left(M_{Pl}\frac{V'(\phi)}{V(\phi)}\right)^2\equiv\lambda^{*2}<2.
\end{equation}
Obviously,   $\mathit{w}_\phi\equiv\frac{P_\phi}{\rho_\phi}=\frac{\dot\phi/2-V(\phi)}{\dot\phi/2+V(\phi)}$, leads to different dynamics for the different potentials.

The axion potential gives the thawing solution, where the field and its corresponding equation of state are almost constant in the early cosmological era, with $\mathit{w}_\phi=-1$, and then starts to evolve after the mass drops below the Hubble parameter, leading to $\mathit{w}_\phi\ge-1$ \cite{Frieman:1995,Caldwell:2005tm}. The axion shift symmetry might allow to tame loop corrections. The condition (\ref{Quintessencecondition}) reads then {\footnote{Note, that for the cosmological application, we have taken, as in \cite{Frieman:1995}, the potential to be $V(\phi)=\mu^4\left(1 + \cos\left(\frac{\phi}{f_a}\right)\right)$. This corresponds to a shift of the minimum in	(\ref{Axionpotential}) with no consequence for the analysis performed in section \ref{subsection-Axion}.}}
\begin{equation}
\label{axionquintessence}
\sin^2\left(\frac{\phi}{f_a}\right)<2\frac{f_a^2}{M_{Pl}^2}\left(1+\cos\left(\frac{\phi}{f_a}\right)\right)^2.
\end{equation}
The requirement $f_a\le M_{Pl}$ allows the axion-like fifth force to be stronger than gravity when $\phi$ gets sufficiently close to 0 for eq. (\ref{axionquintessence}) to be realized. Observational constraints allow this model to be used for quintessence with $\mathit{w_0}\in\left]-1, -0.7\right[$,  $\mathit{w_0}$ being today's value.

The  power law potential gives rise to the tracking solution\cite{Zlatev:1998tr,Steinhardt:1999nw}. This allows for a cosmic evolution from the so-called scaling fixed point $(x,y)=\left(\sqrt{\frac{3}{2}}\frac{1+\mathit{w_m}}{\lambda}, \sqrt{\frac{3}{2}\frac{1-\mathit{w_m}^2}{\lambda^2}}\right)$, with  $x=\frac{\dot{\phi}}{\sqrt{6}M_{Pl}H}$ and $y=\frac{\sqrt{V(\phi)}}{\sqrt{3}M_{Pl}H}$, where matter dominates, to the fixed point $(x,y)=(\lambda^*/\sqrt 6,\sqrt{1-\lambda^{*2}/6})$, where the cosmic acceleration can be realized \cite{Copeland:1997et}. The behaviour of the equation of state is opposite to the previous case, as $w$ slowly decreases with the evolution. Equation (\ref{Quintessencecondition}) gives 
\begin{equation}
\phi^2>\frac{1}{2}p^2M_{Pl}^2,
\end{equation}

Unless the $p$ parameter is tuned to be very small, this calls for trans-Planckian values of the field, as we should have expected since the potential is monotonically decreasing to reach its asymptotic value $V=0$ at infinity. Together with our constraint of weak gravity  $\phi^2\le\frac{(p+2)(2p+1)}{3}M_{Pl}^2$, this leads to:

\begin{equation}
\frac{(p+2)(2p+1)}{3}>\frac{p^2}{2},   
\label{pConstraint}
\end{equation}
which is valid for all positive powers. Of course, the applicability of the effective field theory treatment at trans-Planckian scales is for the least questionable. Observations have led to constrain the tracker equation of state so tightly that the current accepted range of value for the exponent $p$ is very restricted. Indeed, the upper bound on $p$ was argued to be $p<0.107$ in \cite{Chiba:2012cb}, or $p<0.17$ in \cite{Kase:2018aps}, so that positive integers should be excluded, making it difficult to realize power law potentials within the observational bounds in particle physics models.  

The single exponential potential is popular as the cosmological evolution is there described by a closed system of equation \cite{Copeland:1997et,Ferreira:1997hj}. However, the fact that $\lambda^*$ is constant in this case leads to strongly constrain this potential. It is realized again in the fixed point mentioned above but to be reached from the trivial fixed point $(x,y)=(0,0)$ \cite{Tsujikawa:2013fta}. In particular, the transition from the more interesting scaling fixed point $(x,y)=\left(\sqrt{\frac{3}{2}}\frac{1+\mathit{w_m}}{\lambda}, \sqrt{\frac{3}{2}\frac{1-\mathit{w_m}^2}{\lambda^2}}\right)$ is forbidden. This can be circumvented by taking the case of a double exponential potential, as in eq. (\ref{Doubleexponential}). The solution which is realized in this case is a tracking one with constant $\Omega_\phi$\cite{Barreiro:1999zs}.

The exponential potentials with decay constants respecting the upper bound discussed may well fit into the proposed inequality with 

\begin{equation}
\lambda^2\frac{M_{Pl}^2}{f^2}>\frac{3}{2}.	
\end{equation}
For the epoch of cosmic acceleration to be realized we need instead

\begin{equation}
\lambda^2\frac{M_{Pl}^2}{f^2}<2.
\end{equation}
As we see, this seems to leave a window for both the weakness of gravity and the period of cosmic accelerated expansion to be realized through an exponential potential. 

These type of potentials have also been constrained with current observations in the interest of other swampland conjectures, namely the de Sitter and the TCC conjectures \cite{Agrawal:2018own,Bedroya:2019snp}. It was argued in \cite{Agrawal:2018own} we should have for an exponential potential $\lambda^*=\lambda\frac{M_{Pl}}{f_a}\le 0.6$. This was devised to be in agreement with the de Sitter conjecture with the constant $c$ there appearing bounded to be $c\le0.6$. This bound is sensitive to uncertainties in the data as was investigated in e.g. \cite{Heisenberg:2018yae,Akrami:2018ylq}. 
This seems to leave as the only viable conclusion that an exponential quintessence model can only lead to fifth force interactions weaker than gravity. However, \cite{vandeBruck:2019vzd} has hinted to the possibility that dark matter-dark energy coupling may relax constraints on $\lambda$. 

A double exponential is usually devised to respect both constraints coming from big-bang nucleosynthesis and cosmic acceleration. As such, one exponent, $\lambda_1$, is taken to give $\lambda_1\frac{M_{Pl}}{f}\sim 1-10$, while the second is expected to take over at late times and respects the same bounds as those for the single exponent \cite{Agrawal:2018own,Chiba:2012cb,Barreiro:1999zs}. In this case, the weak gravity may be realized in the early Universe as long as the double exponential is concerned, but at late time, one faces the same strong constraints as discussed above.


\section{Multiple Scalar and Moduli Fields}	

We consider now more complex situations with multiple scalar fields. The preeminence of the scalar interaction over the gravitational one has to be formulated in more general terms to account for these cases. In particular, we need to specify what are the processes we should consider to compare scalar and gravitational interactions.

In the case of multiple scalars, we assume that {\it in the appropriate low energy limit, for the leading interaction, the gravitational contribution must be sub-leading}. The focus on the leading scalar interaction can be seen as parallel to constraining the biggest ratio $q/m$ in the WGC.

Let's illustrate the meaning of this statement. First consider the case of a massive scalar $X$, taken to be complex for simplicity. The leading interaction is given by  the Yukawa coupling to another real scalar field $\phi$ and is described by :
\begin{equation}
\mathcal{L}_{int}=\mu \phi |X|^2 + \cdots
\end{equation}
where the dots stand for sub-leading higher order terms. We can write the potential as:
\begin{equation}
 V(X,\phi)= m_X^2 (\phi) \, \,  |X|^2, \qquad{\mu = \partial_\phi m^2_X}
\label{VXtrilineaire}
\end{equation}
The preeminence of scalar interactions must be taken at the mass scale $\sim 2 m_X$ and reads then:
\begin{equation}
\lvert \partial_\phi m_X \rvert  \ge \frac{m_X}{{\tilde M_{Pl}}}    
\label{Dominant1}
\end{equation}
We can square the above three-point amplitudes on each side, $2X \rightarrow \phi$ on the left  and $2X \rightarrow G$, on the right side, where $G$ is the graviton. The comparison concerns then  two  $X X^* \rightarrow X X^*$ processes, at the energy scale $m_X$, one through scalar and the other through graviton exchange. This leads to the following potentials for $X$:
\begin{equation}
V_{scalar}(r)=-\frac{\mu^2}{4m_X^2r}, \qquad{V_{grav}(r)=-\frac{m_X^2}{\tilde M_{Pl}^2r}}
\label{PotentialCompare}
\end{equation}
Now, both scalar and gravitational interactions have similar dependence in the inter-particles distance and the comparison is straightforward:
\begin{equation}
\frac{\mu^2}{4m_X^2}    \ge \frac{m_X^2}{{\tilde M_{Pl}}^2}
\label{MasslessExchange}
\end{equation}
which can be written:
\begin{equation}
\partial_\phi m_X \partial_\phi m_X  \ge \frac{m_X^2}{{\tilde M_{Pl}}^2}    
\label{MasslessExchangeConj}
\end{equation}
In the extremal case saturating the above inequality, the solution is given by:
\begin{equation}
 m^2_X (\phi)= m_0^2 \, \, e^{\pm{ 2 \phi}/{{\tilde M_{Pl}}}}. 
\label{MasslessExchangeConjRes}
\end{equation} 
This is the Swampland Distance Conjecture (SDC) \cite{Ooguri:2006in,Klaewer:2016kiy,Bedroya:2019snp,Grimm:2018ohb,Heidenreich:2018kpg,Lee:2019tst,Brahma:2019mdd}. The inequality (\ref{MasslessExchangeConj}) has been proposed by \cite{Palti:2017elp} in order to retrieve (\ref{MasslessExchangeConjRes}) and  discussed by \cite{Palti:2017elp,Shirai:2019tgr,Gendler:2020dfp} with different motivations.

Let us now move forward and consider another case: a massless complex modulus field $\Phi$, therefore with vanishing potential. We assume again that the theory contains at least one complex scalar field $X$  such that the mass of $X$ and its different couplings are functions of $\phi$. For simplicity, we also assume that $X$ has no tadpole and its vacuum expectation value vanishes, $\langle X \rangle = 0$. Under these assumptions, the scalar potential then takes the form:
\begin{equation}
 V(X,\Phi)= m_X^2 (\Phi) |X|^2 + \cdots \qquad 
m_X^2= m_{X0}^2 + \lambda_\Phi \lvert \Phi \rvert^2 + \cdots
\label{VX1}
\end{equation}
where
\begin{equation}
 \lambda_\Phi = \partial_{\Phi}\partial_{\bar{\Phi}}m^2_X (\Phi, {\bar{\Phi}})
\label{lambdaPhi}
\end{equation}
represents now the leading non-gravitational interaction of $\Phi$. Here, $m_{X0}^2$ is a contribution to the squared mass independent of $\Phi$, but depending on other fields while $\lambda_\Phi$ gives a scalar four-point interaction term of $\Phi$ and $X$  obtained by expanding (\ref{VX1}) in powers of $\Phi$ and $\bar{\Phi}$. The weakness of gravitational interaction becomes a statement comparing on one side the annihilation of two $X$ states into two $\Phi$ state (and vice-versa) and on the other side the same channel through  graviton exchange, both taken at the threshold energy scale  $\sim 2m_X$. 

As the modulus is massless, the gravitational interaction gets an enhancing factor of 2 compared to the massive case, analogous to the case of the gravitational deflection of light. In this case, the statement that the gravitational interaction is weaker reads{\footnote{For real fields, the inequality reads $g^{ij}\partial _i \partial_j  m_X^2 \ge 2 n m_X^2/{{\tilde M_{Pl}}^2}$ where $g^{ij}$ and $n$ are the metric in the space and the number of moduli fields.  The dots in (\ref{VX1}) include $\Phi^2$ and $\bar\Phi^2$ as required to recover the case of real fields scattering and account for an extra factor of $2$.}}:
\begin{equation}
 \partial_{\Phi}\partial_{\bar{\Phi}}m^2_X \ge 2 \frac{m_X^2}{{\tilde M_{Pl}}^2}
 \label{conjectureX}
\end{equation}
If the state $X$ has a self-quartic interaction, then we will also have to check a  similar constraint on the self coupling $\lvert {\tilde \lambda_4} \rvert {\tilde M_{Pl}}^2 \ge m_X^2$.

The extremal case corresponds to the case of equality in (\ref{conjectureX}). It is solved for\footnote{Note that this is not the most general solution but we focus on reproducing the toroidal compactification dependence. Moreover, as the potential \eqref{VX1} and the equation \eqref{conjectureX} are symmetric under the exchange of the real and imaginary part, we choose to focus on the real part of the field only i.e. KK and winding excitation along one of the torus dimensions. }:
\begin{equation}
 m_X^2(\Phi, \bar{\Phi})= m_{-}^2 e^{-\sqrt{2}\frac{\Phi+\bar{\Phi}}{{\tilde M_{Pl}}}} + m_{+}^2 e^{\sqrt{2}\frac{\Phi+\bar{\Phi}}{{\tilde M_{Pl}}}}
\label{KK-winding}
\end{equation}
We can use the following parametrization: 
\begin{equation}
 \Phi = \frac{1}{ \sqrt{2} } (\phi +i \chi), \qquad  e^{\sqrt{2}\frac{\Phi+\bar{\Phi}} {{\tilde M_{Pl}}} } = e^{2 \frac{\phi}{{\tilde M_{Pl}}}}, \qquad   {\rm and} \qquad e^{ \frac{\phi}{{\tilde M_{Pl}}}}=R
 \label{conjectureXpaarametrisation}
\end{equation}
then:
\begin{equation}
 m_X^2(R)= \frac{m_{-}^2}{R^2}  + m_{+}^2 {R^2}
 \label{KK-winding}
 \end{equation}
which is the well known formula  for string states squared masses with the $\frac{m_{-}^2}{R^2}$ as the low energy Kaluza-Klein modes and $m_{+}^2 {R^2}$ the winding modes that are typical to extended objects, strings, winding around a compactified dimension. The  (\ref{KK-winding}) differs sensibly from (\ref{Gonzalo-Ibanez-1}) as it extremizes a different inequality.

Note that in the statement about the preeminence of the scalar interaction, the two fields $\Phi$ and $X$  play a symmetric role.

Now, consider the case where the field $\phi$ is a modulus appearing only as a parameter in the couplings of the massive scalar $X$ ($\langle X \rangle = 0$),  through 
\begin{equation}
 V(X,\phi)= m_X^2 (\phi) X^2  + \sum_{n\ge 4}  \lambda_n(\phi) X^n
\label{VX}
\end{equation}
Then, the condition (\ref{Dominant1}) can be written as:
\begin{equation}
 \frac{|\partial_\phi V(X,\phi)|}{V} \bigg \rvert_{X=0} \ge \frac {\sqrt{\tilde c}}{ M_{Pl}}
\label{VX-1}
\end{equation}
while the condition (\ref{conjectureX}) reads now:
\begin{equation}
 \frac{|\partial_\phi \partial_{\bar {\phi}}V(X,\phi)|}{V} \bigg \rvert_{X=0} \ge \frac { 2 \tilde c}{ M_{Pl}^2}
\label{VX-2}
\end{equation}
where we note the similarity with the Refined de Sitter Conjectures \cite{Obied:2018sgi,Garg:2018reu,Dvali:2018fqu,Andriot:2018wzk,Murayama:2018lie,Ooguri:2018wrx,Dvali:2018jhn,Buratti:2018onj,Roupec:2018mbn,Conlon:2018eyr} (in (\ref{VX-2}) when the second derivative is negative).

A popular way to look at the Weak Gravity Conjecture rests on the fact that the equality in (\ref{WGCcharge-mass-ratio}) relates to the BPS states relation. In \cite{Palti:2017elp}, it was suggested that the  identity satisfied by the central charge in $\mathcal N=2$ supersymmetry \cite{Ferrara:1997tw} 
\begin{equation}
g^{i{\bar j}}D_i{\bar{D}_{\bar j}}|Z|^2 = g^{i{\bar j}}D_i Z {\bar{D}_{\bar j}}{\bar Z} + n |Z|^2
\label{ZforN=2}
\end{equation}
can be used to extract a bound on the mass $m$ as in the BPS case $|Z|=m$:
\begin{equation}
g^{ij}\partial _i \partial_j  m^2 \ge g^{ij}\partial _i m \partial_j m + n m^2
\end{equation}
with derivatives are with respect to scalar fields, and $g_{ij}$ is the corresponding metric. Here, we would like to contemplate a different possibility. Following \cite{Ferrara:1997tw}, the right hand side of (\ref{ZforN=2}) is identified with the scalar potential of the black hole solution, and it was shown that it implies that at the critical point the potential satisfies (in reduced Planck mass units):
\begin{equation}
\partial_i  \partial_{\bar j}   V\bigg \rvert_{critical}= 2 G_{i\bar j} V_{critical}
\label{VforN=2}
\end{equation}
We would like to contemplate here the possibility to extend this relation, beyond its derivation in the $\mathcal N=2$ world, to
\begin{equation}
\lvert  \partial_i \partial_{\bar j} V \rvert \ge c V
\label{VforNnot2}
\end{equation}
as given by (\ref{VX-2}).  Along this line of thought, we note the similarity of (\ref{conjectureX}), up to a factor 2 due to the masslessness of our field $\Phi$, and the equation \cite{Ferrara:1997tw}:
\begin{equation}
  \partial_i \partial_{\bar j} m(\Phi,{\bar \Phi}, p, q) \bigg\rvert_{critical} = \frac {1}{2} G_{i\bar j} (\Phi,{\bar \Phi}) \, \, m(\Phi,{\bar \Phi}, p, q)_{critical}
\label{VforNnot2}
\end{equation}
where $\Phi,{\bar \Phi}$ are moduli fields, $p,q$ electrical and magnetic charges, $m$ is the black hole mass and $G_{i\bar j}$ is the scalar metric on the moduli space.

Finally, let us comment that while supersymmetry was not explicitly invoked here, it might be required to insure the stability of some flat directions, therefore moduli fields, when radiative corrections are taken into account.



\section{Conclusions}

In contrast with the WGC, there is no obvious, no totally convincing road towards uncovering a law governing the scalar potential in quantum gravity. The main ideas have been reviewed in the introduction. Their variety can be considered as an evidence both for the difficulty and risks in writing such constraints and for the interest in investigating their implications.

We postulate that in the appropriate low energy limit, for the leading interaction, the gravitational contribution must be sub-leading. Such a statement is hollow if one does not specify which process is concerned and the energy scale at which the interaction strengths are compared. We provided answers for these questions for some cases and found that we retrieve some forms of the Swampland conjectures.  

The constraint  (\ref{conjecture}) differs from previous proposed inequalities. Indeed, \cite{Palti:2017elp, Heidenreich:2019zkl} focused on massless scalars and their role in the formation of gravitational bound states. Strictly speaking, the logic behind their inequalities would lead to (\ref{MasslessExchangeConj}) but with an opposite sign for the r.h.s. part. This is due to the fact that their arguments constrain repulsive interactions to be stronger than gravitational one, while the scalar mediated one is attractive. While the logic in this work differs, in the massless case  (\ref{MasslessExchange}) agrees with one of the proposals of \cite{Palti:2017elp}, that was also discussed further in \cite{Gonzalo:2019gjp,Shirai:2019tgr,Gendler:2020dfp}. This is all but surprising as the different arguments were put such as one recovers the SDC, which corresponds to the ubiquitous Kaluza-Klein states present in String theory compactifications. Our analysis differs also in the fact that we have also considered self-interacting scalars but only focused on the case of neutral states.

The conjecture presented in \cite{Gonzalo:2019gjp} leads to an inequality  that would constrain in qualitatively similar manner attractive self-interactions for a massive particle (non-tachyonic), but we were not able to recover their coefficients for the different contributions. Moreover, the field dependence of the extremal states squared mass (\ref{Gonzalo-Ibanez-2}) differs sensibly from our result (\ref{KK-winding}). 

The main playground for testing different conjectures about quantum gravity is string compactifications and their effective supergravity theories. While they represent an opportunity to put the conjecture on firm grounds (see \cite{Cecotti:2020rjq} for a recent proposal), one should be able to disentangle what is due to generic quantum gravity from what is due to supersymmetry, other symmetries or just consistency of the precise string theory compactification. Here, we have kept the analysis on a very basic level which we believe is sufficient to stress the main points.  We plan to test our constraints in string compactification models in the future.

We end by mentioning two immediate remarks. For the Standard Model Higgs scalar, it was found that the running quartic coupling vanishes at  energies of order $10^{11}$ GeV \cite{Degrassi:2012ry}, we should therefore contemplate this intermediate energy scale as an ultra-violet cut-off.  Scalar interactions determine the behaviour of  spherically symmetric cosmological clumps. The size and dynamics of these objects is different depending on the quartic self-interaction coupling $\lambda$. For the case of repulsive complex scalars, massive boson stars, with masses comparable to the fermionic ones, are allowed only when the relevant {\it relativistic} parameter $\lambda M_{Pl}^2/m^2$ is big \cite{Colpi:1986ye}. This is a prediction of the weak gravity conjecture discussed here. 

Going through the implications of our weak gravity requirement we recovered, in the corresponding cases and forms, some of the Swampland program expectations: the Axion Weak Gravity Conjecture, the Swampland Distance Conjecture, the string Kaluza-Klein and winding modes mass formula and the Swampland de Sitter Conjecture. It would be interesting to investigate if a  formulation from general principles of the preeminence of scalar interactions when compared to gravitational ones can lead to a unified Swampland conjecture that rules them all.

\noindent
\section*{Acknowledgments}
We acknowledge the support of  the Agence Nationale de Recherche under grant ANR-15-CE31-0002 ``HiggsAutomator''.


\noindent

\providecommand{\href}[2]{#2}\begingroup\raggedright\endgroup

\end{document}